\documentclass[12pt]{article}
\usepackage{epsfig}
\def\lsim{\raise0.3ex\hbox{$<$\kern-0.75em\raise-1.1ex\hbox{$\sim$}}}
\def\gsim{\raise0.3ex\hbox{$>$\kern-0.75em\raise-1.1ex\hbox{$\sim$}}}
\def\noi{\noindent}
\def\nn{\nonumber}
\def\bea{\begin{eqnarray}}  \def\eea{\end{eqnarray}}
\def\beq{\begin{equation}}   \def\eeq{\end{equation}}

\def\beeq{\begin{eqnarray}} \def\eeeq{\end{eqnarray}}

\begin{document}
\vbox to 2 truecm {}
\begin{center}
{\large \bf NEXT-TO-LEADING ORDER DETERMINATION OF}
\vskip 3 truemm
{\large \bf  FRAGMENTATION FUNCTIONS} \\  \vskip 1 truecm
{\large L.~Bourhis$^{3}$, M.~Fontannaz$^2$, J.Ph.~Guillet$^1$, M.~Werlen$^1$} \\ 

\begin{enumerate}
\item {\em Laboratoire de Physique Th\'eorique LAPTH,
\footnote{UMR 5108 du
CNRS, associ\'ee \`a l'Universit\'e de Savoie.}\\
B.P. 110, F-74941 Annecy-le-Vieux Cedex, France} \\
\item {\em Laboratoire de Physique Th\'eorique
\footnote{Unit\'e mixte de recherche (CNRS) UMR 8627}\\
Universit\'e de Paris XI, B\^atiment 210,
F-91405 Orsay Cedex, France}\\
\item {\em Department of Theoretical Physics \\
South Road, Durham City DH1 3LE, United Kingdom}
\end{enumerate}
\end{center}

\vskip 1 truecm
\begin{abstract}
We analyse LEP and PETRA data on single inclusive charged hadron cross-sections to establish new sets
of Next-to-Leading order Fragmentation Functions. Data on hadro-production of large-$p_{\bot}$ hadrons
are also used to constrain the gluon Fragmentation Function. We carry out a critical
com\-pa\-ri\-son with other NLO parametrizations. \end{abstract}

\vskip 3 truecm
\rightline{hep-ph/0009101}
\rightline{Durham 00/28}
\rightline{LAPTH-802/00}
\rightline{LPT-Orsay/99/94}
\rightline{September 2000}
\newpage
\pagestyle{plain}
\baselineskip=20 pt

\section{Introduction}
\hspace{\parindent} Inclusive hadron production in various processes gives the possibility to carry
out quantitative tests of the QCD-improved parton model. The essential property of QCD
cross-sections which allows comparisons between various processes is the factorization property.
Indeed thanks to the factorization theorem \cite{1r}, the cross-sections are written as
convolutions of basic building blocks, such as the quark and gluon distributions in the incoming
hadrons, the hard subprocesses describing the large-angle scattering of partons, and the
fragmentation functions of quarks and gluons into hadrons. The parton distributions and
fragmentation functions can be defined in a universal way. Therefore these distributions measured
in one reaction can be used to perform predictions for another reaction. On the other hand, the
subprocess cross-section is entirely calculable in perturbative QCD, the only free parameter being
$\Lambda_{QCD}$. \par

Many processes involving fragmentation functions can be related in this manner. Among them the best
studied are the $e^+e^-$-annihilation into hadrons $e^+e^- \to hX$, the observation of hadrons in DIS
experiments $\ell p \to \ell hX$, and the hadro- and photo-production of large-$p_{\bot}$ hadrons
$h_ah_b \to h_cX$ and $\gamma h_a \to h_cX$. Clearly quantitative studies of these reactions can only
be performed if sets of next-to-leading order (NLO) fragmentation functions are available. \par

While there are several sets of parton distribution functions for the proton and the pion, mainly
extracted from DIS data, few sets of fragmentation functions exist. It is only recently that many
precise data on the hadron energy spectrum in $e^+e^-$-annihilation became available from LEP
experiments, completing those obtained at lower energy at DESY, SLAC and TRISTAN. The accuracy of
these data is remarkable~; they should allow a determination of fragmentation functions with
a precision which approaches that obtained in the parton distribution functions. A first
step towards a NLO analysis of fragmentation functions has been done by Chiappetta et al.
(CGGRW collaboration) \cite{2r}, who studied the fragmentation into $\pi^0$. As LEP data
did not exist at that time, these authors also made a detailed study of large-$p_{\bot}$
$\pi^0$ cross-sections, measured by ISR experiments and the UA2 collaboration, in order to
constrain their parametrizations. LEP data was used by Binnewies, Kniehl and Kramer
(BKK coll.), in association with lower energy results of PEP and PETRA, to determine the
parton fragmentation into charged pions and kaons, and into neutral kaons \cite{3r,4r}.
From these results and the measured ratio $p/\pi$, they indirectly obtained 
fragmentation functions for charged particles\footnote{During the completion of this work, 
two new papers on fragmentation 
functions apppeared\cite{kkp,fra2}. We
shall briefely comment on them.}. \par

We pursue these studies and propose parametrizations\footnote{These fragmentation functions are
available on request (guillet@lapp.in2p3.fr).} 
of the parton fragmentation
functions into charged particles by directly analyzing the cor\-res\-pon\-ding $e^+e^-$ data. The
interest of such functions is that they can be used in all reactions in which there is no particle
identification. On the other hand this set of fragmentation functions can be compared with the BKK
set, thus offering an estimation of the ``theoretical error'' embedded in the parametrizations and
which comes from various theoretical assumptions used to perform fits to data. We shall
see, in section 3, that the gluon fragmentation function is poorly constrained by $e^+e^-$
data. Therefore we shall complete our anlysis by that of the hadron-production of
large-$p_{\bot}$ charged hadrons, a reaction which is sensitive to the gluon fragmentation
function. \par

This paper is organized as follows. In section 2, we present the theoretical expressions we use to
calculate the single inclusive hadron cross-section in $e^+e^-$-annihilation, and we describe the
parametrization of the non-perturbative boun\-da\-ry conditions associated to the DGLAP evolution
equations for the fragmentation functions. We also pay attention to an improved choice of the
renormalization and factorization scales. In section 3, we describe the fits to LEP and PETRA data
and discuss our results. Then we study, in section 4, the constraints put on the gluon
fragmentation function by large-$p_{\bot}$ hadron production. We conclude in section 5.

\section{Single Inclusive Cross Section in e$^+$e$^-$\break \noindent Annihilation}
\hspace{\parindent} In the QCD-improved parton model, the single inclusive cross-section is given by
the expression

\beq
\label{1e}
{dN^h \over dz} = {1 \over \sigma_{tot}} \ {d\sigma (e^+e^- \to hX) \over dz} = \sum_{a=q,\bar{q},g}
\int_z^1 {dy \over y} D_a^h\left ( y,M^2 \right ) {1 \over \sigma_{tot}} \ {d\sigma_a \over dv}
\left ( v = {z \over y}, \mu^2, M^2, Q^2 \right ) \ . \eeq

\noi The longitudinal variables $z = \displaystyle{{2p_h\cdot Q \over Q^2}}$ and $v =
\displaystyle{{2p_a \cdot Q \over Q^2}}$ measure the fraction of the energy $\sqrt{Q^2}$, available in
the $e^+e^-$ CMS, carried away by the hadron $h$ and the parton $a$. The fragmentation functions
$D_a(u,M^2)$ describe the non-perturbative transition of parton $a$ into hadron $h$. Only the
dependence on the factorization scale $M$ is perturbatively calculable and given by the DGLAP
evolution equation

\beq
\label{2e}
M^2 \ {\partial D_a^h(M^2) \over \partial M^2} = \sum_a P_{ba}^T \otimes D_b^h(M^2)
\eeq

\noi where $\otimes$ indicates a convolution~: $(f \otimes g)(x) = \int_0^1 dy \ dv\ f(y) \
g(v) \ \delta (yv-x)$. The time-like kernels are expansions in $\alpha_s(M^2)$

\beq
\label{3e}
P_{ba}^T(M^2) = {\alpha_s \over 2 \pi}(M^2) \ P_{ba}^{(0)} + \left ( {\alpha_s \over 2\pi}(M^2) \right
)^2 \ P_{ba}^{T(1)} + \cdots  \eeq

\noi In this paper we work at the NLO approximation and when solving (\ref{2e}), we use the
NLO kernels $P_{ba}^{T(1)}$ \cite{5r,6r}. Eq. (\ref{2e}) can be solved more easily by
working in moment-space. We use this formalism here as well as a computer code written by P.
Nason \cite{7r}. \par

The hard subprocess cross-section $d\sigma_a/dv$ describes the production of a parton $a$ in
$e^+e^-$-annihilation. (We consider only the sum of the transverse and longitudinal cross sections). It is given by an expansion in $\alpha_s(\mu^2)$ where $\mu^2$ is the
renormalization scale~:

\beq
\label{4e}
{d\sigma_a \over dv} (v, \mu^2, M^2, Q^2) = {d\sigma_a^{(0)} \over dv}(v, Q^2) + {\alpha_s \over 2
\pi} (\mu^2)   {d\sigma_a^{(1)} \over dv} (v, M^2,Q^2) + \cdots 
 \eeq

\noi We use expressions calculated at order $\alpha_s(\mu^2)$ \cite{8r}. For the
total cross-section $\sigma_{tot}$ we also use the $O(\alpha_s)$
expression\footnote{Expressions of $\sigma^{Born}$ which also contain the $Z^0$ contribution
may be found in ref. \protect{\cite{9bis,10bis}}.}~: 

\beq
\label{5e}
{1 \over \sigma_{tot}} = {1 \over N_c {4 \pi \alpha^2 \over 3Q^2} \sum\limits_{i=1}^{N_f} e_{q_i}^2
\left ( 1 + {\alpha_s(\mu^2) \over \pi} \right )} \simeq {1 \over \sigma^{Born}} \left ( 1 -
{\alpha_s \over \pi} (\mu^2) \right ) \quad . \eeq

Throughout this paper, we work at NLO accuracy and do not take into account the NNLO
expressions calculated by Rijken and van Neerven \cite{10bis}. Several reasons impose this
choice. First, the full NNLO correction is not known~; the 3-loop DGLAP kernel is missing.
Second, we shall use these fragmentation functions in the calculation of cross section for
which only NLO expressions are available (for instance large-$p_{\bot}$ inclusive cross
sections)~; it appears more coherent to also use NLO parametrizations of the fragmentation
functions. One must also notice that the NLLO corrections to the $e^+e^-$ inclusive cross
section are very small when $z$ is not too close to zero or one \cite{10bis}. In this
paper, we study data in the range $.12 \leq z \ \lsim \ .9$, and for $z \simeq .9$ the
experimental errors are larger than the NLLO corrections. \par

In order to neglect kinematical higher twists of order $m^2/\vec{p}^{\, 2}$ where $m$ is a hadron
mass and $\vec{p}$ its momentum, we restrict the $z$-range ($z \sim 2|\vec{p}|/\sqrt{Q^2})$
studied by the condition $z \ \gsim \ .12$. This condition also allows us to neglect ``MLLA''
effects \cite{10r} which show up at small $z$ and are not included in the NLO formalism
described in this section. \par

On the other hand the ${\cal O}(\alpha_s^2)$ corrections to the longitudinal cross section
$d\sigma_L/dz$ calculated in \cite{10bis} are non-negligible. However we will not use this
cross section to constrain to gluon fragmentation functions $D_g(z,Q^2)$, because data at
large $z$ ($z \ \gsim \ . 4)$ is scarce and not very precise. \par

Finally let us mention that we work in the $\overline{MS}$ scheme and that we use massless
expressions for the DGLAP kernels and the hard cross sections. However we take into account
the bottom threshold by starting the evolution of the $b$-quark fragmentation function at
$M^2 = m_b^2$. \par

The cross-section (\ref{1e}) depends on arbitrary scales which show up in the course of the
theoretical calculations~: the renormalization scale $\mu$ and the factorization scale $M$. They
must be chosen of the order of $\sqrt{Q^2}$, but their precise value cannot be determined by any
fondamental rule. However several approaches exist which propose prescriptions to improve the
choice of these scales~; we adopt here the ``Principle of Minimum Sensitivity'' criterion
\cite{9r}. At the order $\alpha_s(\mu^2)$ at which $d\sigma_a/dv$ is calculated (LO calculation in
$\alpha_s(\mu^2)$), no prescription constrains $\mu$ and we choose $\mu = M$. Then we fix $M$ using
the PMS criterion

\beq
\label{6e}
\left . M^2 \ {\partial \over \partial M^2} \left ( {dN \over dz} \right ) \right
|_{M=M_{opt}} = 0  \quad .\eeq

We find by a numerical study of (\ref{6e}) that $M_{opt}^2$ is much smaller than $Q^2$. In a
large range in $z$, we have $M_{opt} \simeq \sqrt{Q^2}/5$. However at small values of $z$ ($z
\ \lsim \ .2$), $M_{opt}$ starts to increase to values as large as $\sqrt{Q^2}$. At Mark II \cite{MarkII}
energy ($\sqrt{Q^2} = 29$~GeV), the PMS criterion does not work for $z \ \gsim \ .3$~; there is
no solution of equation (\ref{6e}). Because of this result, we shall only study data with
$\sqrt{Q^2} \ \gsim \ 35$~GeV. This leads us to discard PEP data. \par

As already discussed, the non-perturbative physics of the cross-section (\ref{1e}) is entirely
contained in the fragmentation functions $D_a^h (y, M^2)$. The perturbative evolution of
these fragmentation functions is given by eq. (\ref{2e}) which must be supplemented by a
non-perturbative initial condition. The latter is fixed by the $y$-behavior of the
fragmentation functions at an initial scale $Q_0^2$. As we are interested in the use of the
fragmentation functions in reactions where the scale $M$ can be of the order of a few GeV
(hadro- or photoproduction of large-$p_{\bot}$ particles), we start the evolution at a
small value of $M$, namely $M^2 = Q_0^2 = 2$~GeV$^2$ ($M^2 = m_b^2$ for $D_b$). At this
value of $M$, we fix the shape of the fragmentation function by using a simple and standard
parametrization

\bea
\label{7e}
&&D_g\left ( y,Q_0^2 \right ) = N_g (1 - y)^{\beta_g} \ y^{\alpha_g} \nn \\
&&D_u \left ( y,Q_0^2 \right ) = \left ( N_u(1 - y)^{\beta_u}  + \bar{N}_u (1 -
y)^{\bar{\beta}_u} \right ) y^{\alpha_u} \nn \\
&&D_{d+s}\left ( y,Q_0^2 \right ) = \left ( N_{d+s}(1 - y)^{\beta_{d+s}} +
\bar{N}_{d+s} (1 - y)^{\bar{\beta}_{d+s}} \right ) y^{\alpha_u} \nn \\
&&D_c\left ( y,Q_0^2 \right ) = N_c (1 - y)^{\beta_c} \ y^{\alpha_c} \nn \\
&&D_b\left ( y,m_b^2 \right ) = N_b (1 - y)^{\beta_b}  y^{\alpha_b} \quad . \eea

Let us say a few words on the theoretical and ``experimental'' reasons for this
parametrization. First we notice that the $e^+e^-$-annihilation cross-section is only
sensitive to the sum $D_{d+s} = D_d + D_s$. So we are not able to determine $D_d$
and $D_s$ separately. This degeneracy could be lifted by looking at DIS data, or at data on
the production of large-$p_{\bot}$ particles. However the large -$p_{\bot}$ production
cross-sections, at UA1 and UA2 energy, are sensitive to the gluon fragmentation function,
and very little to the $d$-quark fragmentation function. Therefore in this paper, we will
not give separate descriptions of the $d$- and $s$-quark fragmentation functions. On the
other hand, in order to reduce the number of free parameters of the fit, we assume that the
small-$z$ behaviour of the light quarks are the same~: $\alpha_u = \alpha_d = \alpha_s$.
\par

By performing a fit to $e^+e^-$ data (see next section), we observed that several
parameters  of the input (\ref{7e}) are strongly correlated with each other. For instance
$N_g$, $\alpha_g$ and $\beta_g$ are strongly correlated, as well as $N_u$,
$\bar{N}_u$, $\beta_u$, and $\alpha_{u}$, $\bar{N}_{d+s}$ and $N_g$. The
$b$-parameters also are strongly correlated with each other. Because of these
correlations, the fitting procedure is very lengthy and we cannot obtain a
positive definite error matrix. Therefore we proceed in two steps. First we
perform a fit with all parameters free. Then, in order to reduce the
correlations, we fix some parameters to the values obtained in the first fit.
This will be discussed in detail in the next section. \par

We also observed a strong correlation between the gluon parameters and the functional shape
of the input distribution $D_{d+s}(y, Q_0^2)$. A simple form of the type $N(1 -
y)^{\beta} y^{\alpha}$ leads to a gluon fragmentation function which is in disagreement
with large-$p_{\bot}$ charged particle data of UA1 \cite{11r}. Thus we use more flexible shapes, 
as those of formula (\ref{7e}), in order to try to decorrelate, as far as possible, 
gluon and light quark parameters. \par

In the present paper, we do not intend to determine from data the value of
$\Lambda_{\overline{MS}}$. This value might be sensitive to power corrections
\cite{9bis,aleph} and such an analysis is beyond the scope of this work.
Here we use a fixed value of $\Lambda_{\overline{MS}}^{(4)}$~: $\Lambda_{\overline{MS}} =$
300 MeV, in agreement with the CTEQ4M \cite{29r} and MRST \cite{mrs99} distribution functions. 
One must note that this value is quite in agreement
with the new value obtained in Ref\cite{kkp} in a fit to $e^+e^-$ annihilation data.
 
\section{Analysis of e$^+$e$^-$-annihilation data}
\hspace{\parindent} 
Inclusive charged particle data from CELLO \cite{cello} 
at $\sqrt{Q^2}$ =35 GeV, from TASSO \cite{tasso} at $\sqrt{Q^2}$ =44 GeV,
from AMY \cite{amy} at $\sqrt{Q^2}$ =58 GeV, from DELPHI \cite{del94}
and SLD \cite{sld} at $\sqrt{Q^2}$ =91.2 GeV are used in the fits. To constrain 
the flavor dependent parameters,
the uds, c and b flavor enriched samples from ALEPH \cite{aleph}, 
DELPHI \cite{del98} and OPAL \cite{opal98} are  used.
A b enriched sample from DELPHI \cite{del94} has also been included. 
\par
To take into account the experimental systematics errors, avoiding
the treatment of correlated errors, the following
procedure has been adopted: the normalization errors are not
included in the $\chi^2$ evaluation but a normalization for
each experiment is taken as a parameter of the fit allowed
to vary within 3 standard deviations of the quoted experimental uncertainty.
For ALEPH, as we use enriched
samples in the fit, we do not used the all charged data which has correlated errors 
\cite{aleph}. The normalization of ALEPH ligth quark sample is kept fixed to 1.,
while a common normalization $N_{OPAL}$ ($N_{DELPHI\ 94}$, $N_{DELPHI}$) 
is allowed for all OPAL samples (DELPHI 94 samples \cite{del94}, DELPHI \cite{del98}).
\par

Using the program Minuit \cite{mini}, we performed a first fit to data, 
obtaining a reasonable result with $\chi^2 = 215$ for $217$ data points
and $25$ parameters.
In particular the data normalization factors are very close to one.
As discussed in the preceding section, we
observe strong correlations between the parameters and the error matrix is not
positive definite. \par

The strong correlation between $\beta_u$ and $\bar{\beta}_u$, $\beta_{d+s}$ and
$\bar{\beta}_{d+s}$ leads us to fix the diffferences $\bar{\beta}_u - \beta_u$
and $\bar{\beta}_{d+s} - \beta_{d+s}$ to the results of the first fit namely
$\bar{\beta}_u - \beta_u = 1.15$ and $\bar{\beta}_{d+s} - \beta_{d+s} = 3.6$.
Similarly, we fix $\alpha_g = 0$ and $\alpha_{b} = - 1.7$. The $\chi^2$ is now
204 for $217$ points and $14$ parameters. 
But the fit is not yet converging and the
error matrix is not positive definite. The source of this problem lies in the
remaining strong correlation between $N_b$ and $\beta_b$. Fixing also $N_b$ and
$\beta_b$ we eventually get a convergent fit. Relaxing the condition $\alpha_g =
0$ also lead to a convergent fit with a positive definite error matrix and the
following parameters for the non-perturbative inputs at $Q_0^2 = 2$~GeV$^2$. (The
data normalization factors $N_i$ are also fixed to the values obtained in the
first fit)

\begin{equation}
\begin{array}{ll} 
D_g(y,Q_0^2) &= 2.11  (1-y)^{1.37} y^{-.71}\\
D_u(y,Q_0^2) &=  \left  (3.45 (1-y)^{2.25}  - 0.955(1-y)^{3.4} \right ) y^{-0.68}  
\\ D_{d+s}(y,Q_0^2) &=  \left  (0.95(1-y)^{.945}  + 10.43 (1-y)^{4.545} \right )
y^{-0.68}\\ D_c(y,Q_0^2) &= 5.15 (1-y)^{3.87} y^{-0.78}  \\
D_b(y,Q_0^2) &= 0.897 (1-y)^{3.7} y^{-1.7}  \\
N_{ALEPH \ b} &= 1.03\\
N_{ALEPH \ c} &=1.014\\
N_{OPAL} &=0.999\\
N_{DELPHI\ 94} &=1.0006\\
N_{DELPHI}    &=1.0005\\
N_{SLD} &= 0.9998\\
N_{CELLO} &= 1.0055\\
N_{AMY} &=1.0117\\
N_{TASSO} &=0.9835\\
\end{array}
\label{8e}
\end{equation}

The corresponding $\chi^2$ is 201 for $217$ data points and $13$ parameters. \\

      A technical discussion of errors and correlation matrices is given in 
appendix 1. Here we just quote the results of a full analysis 
(MINOS \cite{mini},\cite{eadie})
on the parameter $N_g$. We obtain a variation of $\chi^2$ by one unit when
increasing (decreasing) the value of $N_g$ to 5.25 (1.10), leaving the other
parameteres free. These alternative fits are studied and used in section 4
where we discuss large-pt data.


One notices that the gluon input is flat, comparable to the light quark fragmentation
functions. However $e^+e^-$ inclusive cross sections poorly constrain the gluon fragmentation
function and the gluon parameters are determined with a large errror. The gluon
fragmentation function corresponds to an ${\cal O}(\alpha_s)$ correction to the
Born cross section, and it is quite sensitive to the functional forms of the light
quark inputs. Therefore we cannot  draw any physical conclusion from this result.
However, it turns out that the gluon (\ref{8e})  gives a good description of the
UA1 large-$p_{\bot}$ data. This point will be developed in section~4. \par

It is interesting to compare our fit to the corresponding data by using ratios and linear scales
which exhibit agreements or disagreement more clearly. 
The experimental error is the quadratic sum of statistical and systematic errors.
 
\begin{figure}[htb]
\begin{center}
\epsfig{file=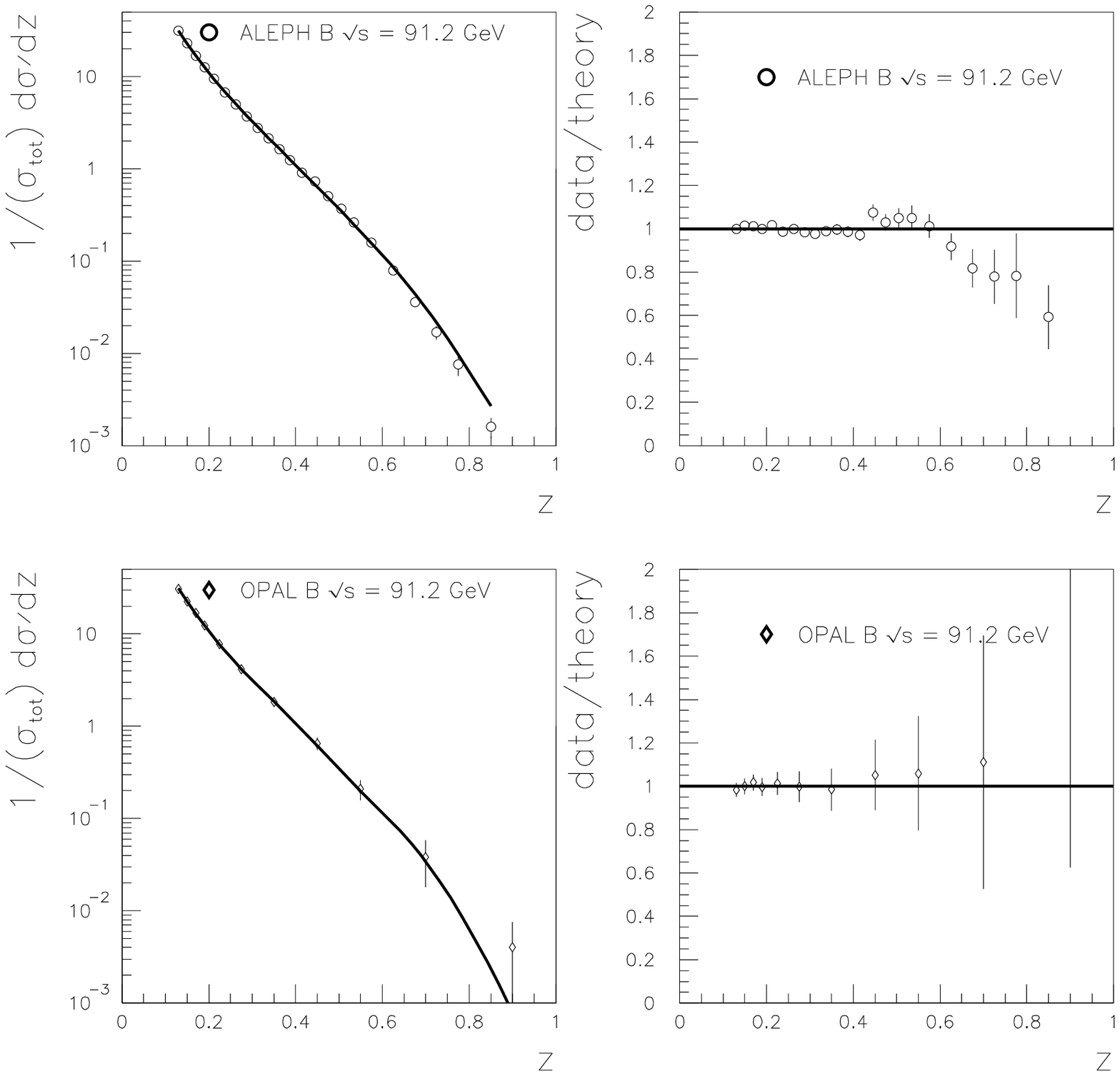,width=14.cm}
\end{center}
\caption{NLO inclusive charged particle production in $e^+e^- \to bX$ collisions
at $\sqrt{s}=91.2$ GeV with optimized scales
and with fragmentation functions obtained 
here  (formula \ref{8e}) compared to data of the ALEPH ~\protect\cite{aleph} 
and OPAL collaboration ~\protect\cite{opal98}. 
}
\label{figb1}
\end{figure}

\begin{figure}[htb]
\begin{center}
\epsfig{file=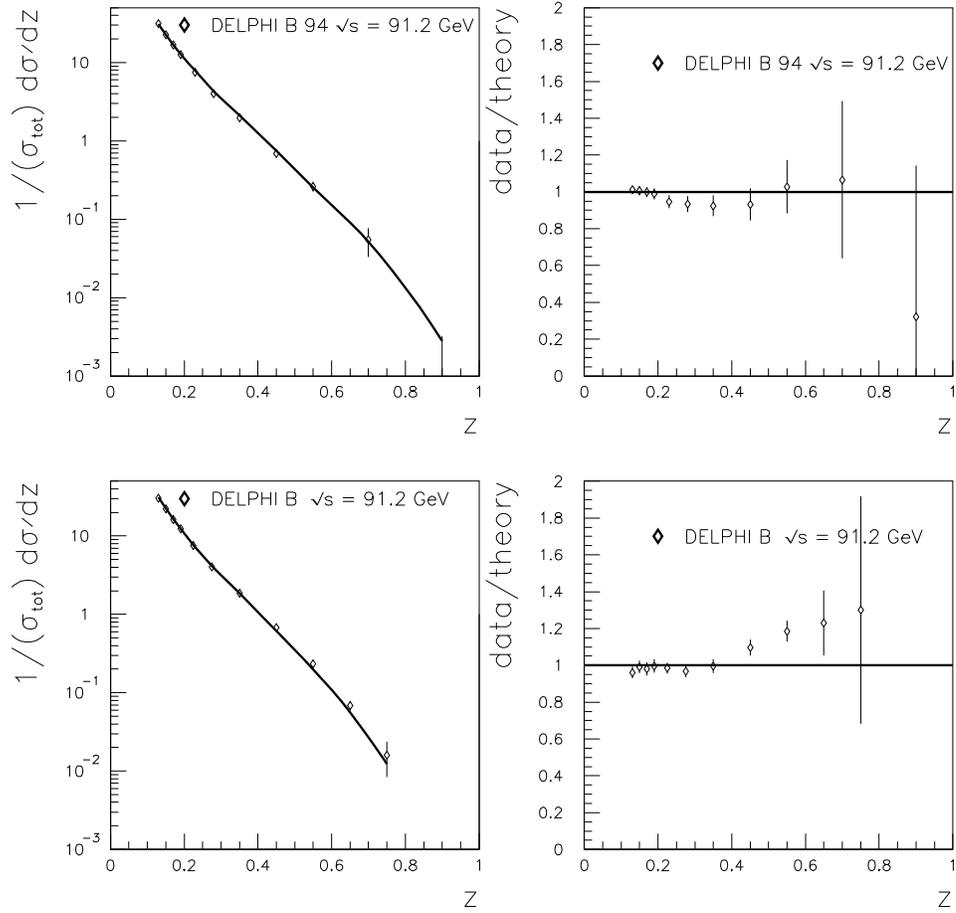,width=14.cm}
\end{center}
\caption{NLO inclusive charged particle production in $e^+e^- \to bX$ collisions
at $\sqrt{s}=91.2$ GeV with optimized scales
and with fragmentation functions obtained 
here (formula \ref{8e}) compared to data of the DELPHI collaboration 
taken in 1994 (top) ~\protect\cite{del94} and in 1991-1993
~\protect\cite{del98}. 
}
\label{figb2}
\end{figure}

\begin{figure}[htb]
\begin{center}
\epsfig{file=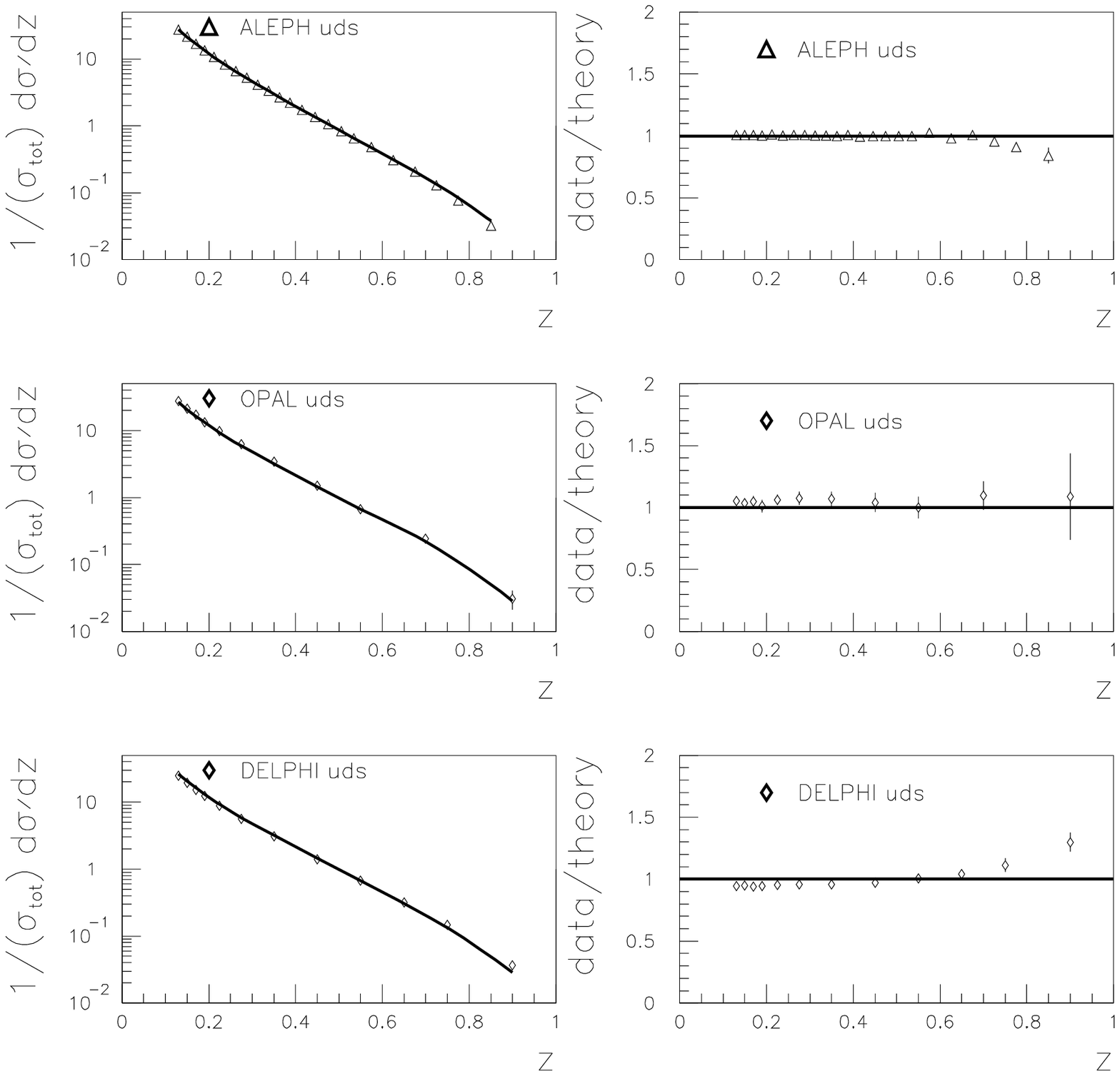,width=14.cm}
\end{center}
\caption{NLO inclusive charged particle production in $e^+e^- \to u,d,sX$ collisions
at $\sqrt{s}=91.2$ GeV with optimized scales
and with fragmentation functions obtained 
here (formula \ref{8e}) compared to data of the ALEPH ~\protect\cite{aleph}, OPAL~\protect\cite{opal98}
and DELPHI ~\protect\cite{del98} collaboration. 
}
\label{figuds}
\end{figure}

\begin{figure}[htb]
\begin{center}
\epsfig{file=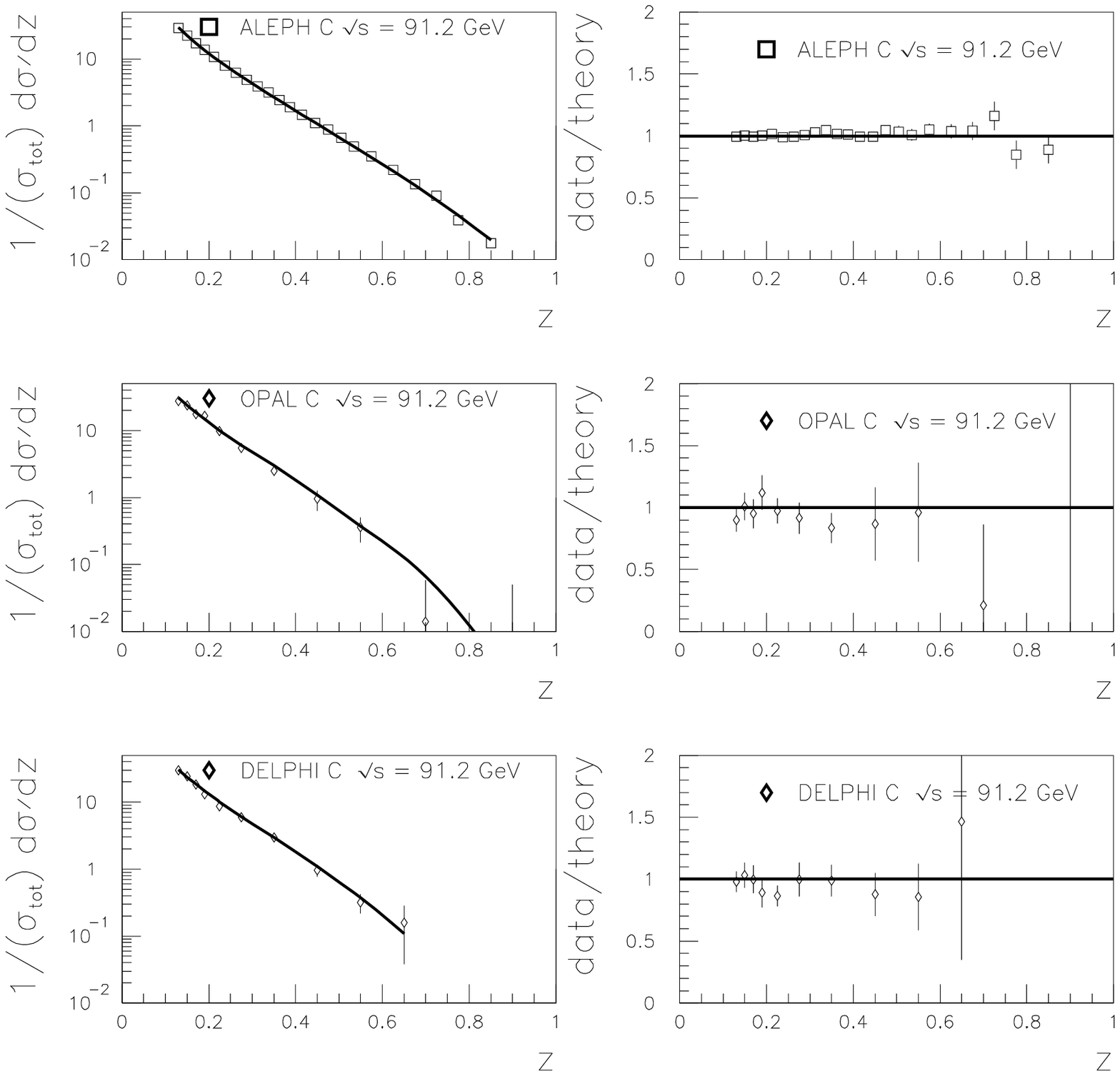,width=14.cm}
\end{center}
\caption{NLO inclusive charged particle production in $e^+e^- \to cX$ collisions
at $\sqrt{s}=91.2$ GeV with optimized scales
and with fragmentation functions obtained 
here (formula \ref{8e}) compared to data of the ALEPH ~\protect\cite{aleph}, OPAL~\protect\cite{opal98}
and DELPHI ~\protect\cite{del98} collaboration. 
}
\label{figc}
\end{figure}

\begin{figure}[htb]
\begin{center}
\epsfig{file=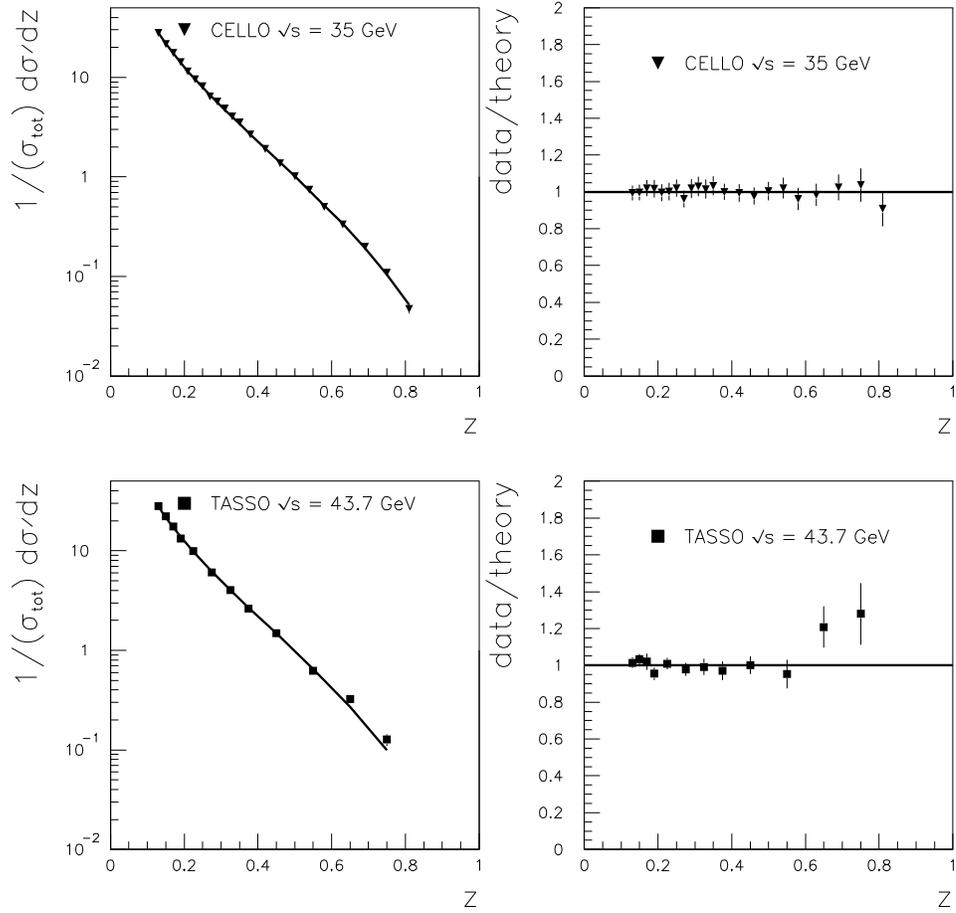,width=14.cm}
\end{center}
\caption{NLO inclusive charged particle production in $e^+e^- \to hX$ collisions
with optimized scales
and with fragmentation functions obtained 
here (formula \ref{8e}) compared to data at $\sqrt{s}=35$ GeV from the CELLO collaboration
~\protect\cite{cello} and at $\sqrt{s}=44$ GeV from the TASSO collaboration
~\protect\cite{tasso}. 
}
\label{figpetra}
\end{figure}

\begin{figure}[htb]
\begin{center}
\epsfig{file=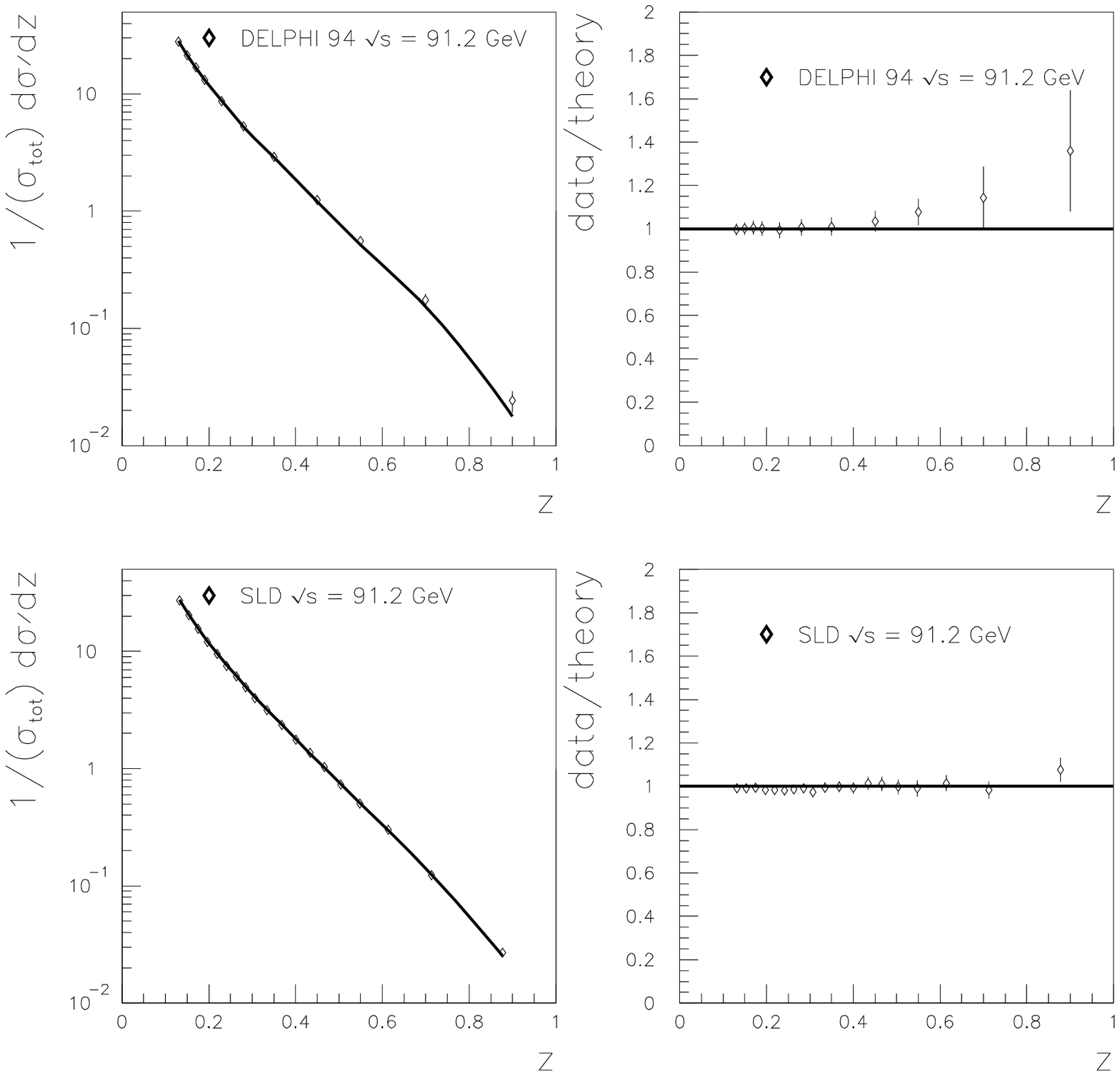,width=14.cm}
\end{center}
\caption{NLO inclusive charged particle production in $e^+e^- \to hX$ collisions
at $\sqrt{s}=91.2$ GeV with optimized scales
and with fragmentation functions obtained 
here (formula \ref{8e}) compared to data of the DELPHI ~\protect\cite{del94} and SLD ~\protect\cite{sld}
collaboration. 
}
\label{figallch}
\end{figure}

Let us start the discussion by the
$b$-enriched cross sections (Fig. \ref{figb1},\ref{figb2}). The agreement is generally good, 
but one notices at large $z$ some
discrepancies between the various data sets which might indicate that the systematic errors are
underestimated. The light and $c$-quark theo\-re\-ti\-cal cross sections and data are displayed in  
Fig. \ref{figuds}, Fig. \ref{figc} and Fig. \ref{figpetra}. CELLO and TASSO data (Fig.
\ref{figpetra}) are mainly sensitive to the $u$ and $c$-quark fragmentation
functions, whereas light-quark enriched LEP sample (ALEPH uds, DELPHI, OPAL uds, DELPHI uds)
are sensitive to the $d$ and $s$-quark fragmentation function. 
Here one also notices some dispersion of the LEP light
quark data at large $z$. The all charged data from DELPHI and SLD, also included in the
fits, are shown in Fig. \ref{figallch}. \par

\begin{figure}[htb]
\begin{center}
\epsfig{file=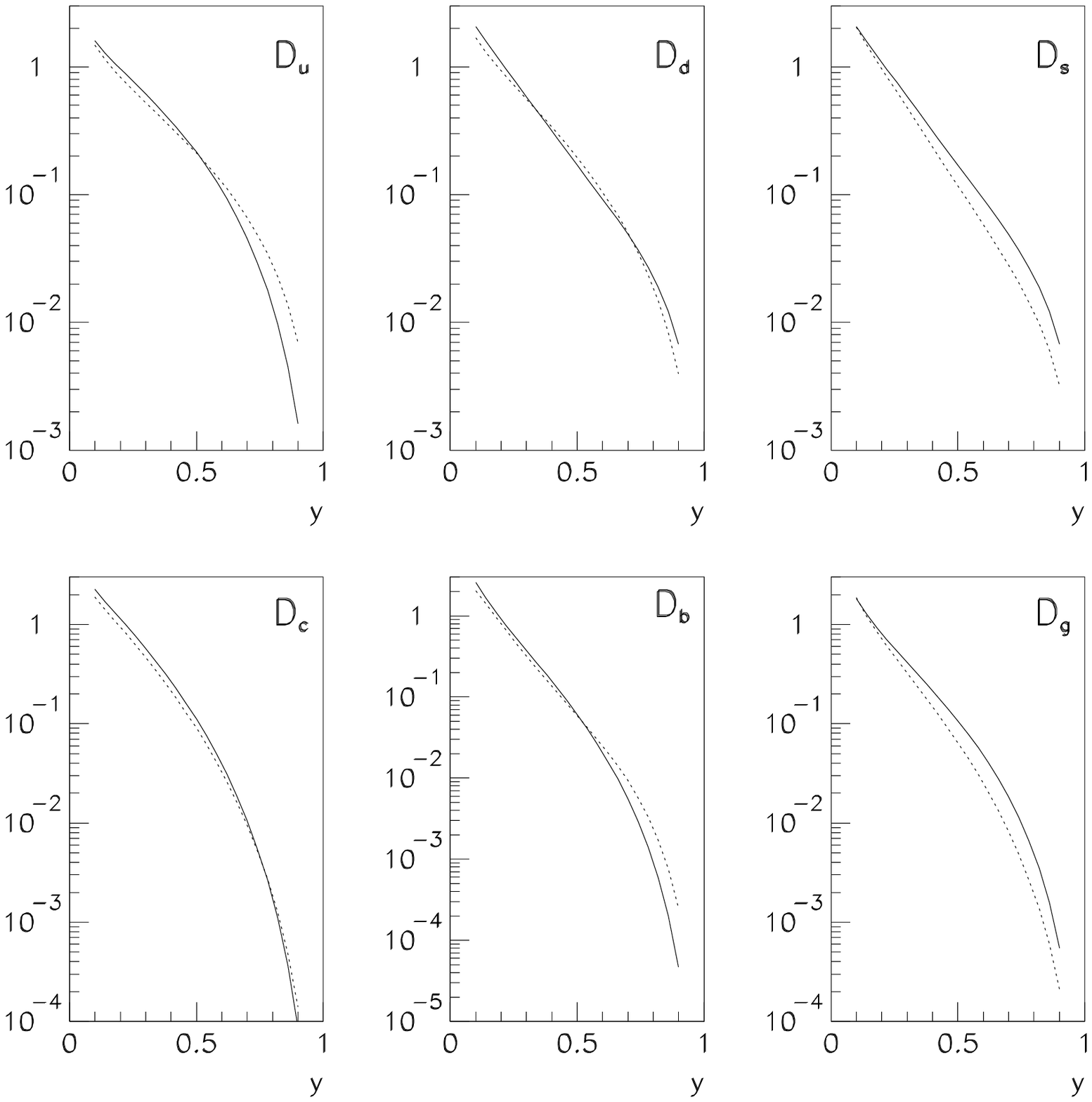,height=14.cm,width=14.cm}
\end{center}
\caption{Quark and gluon fragmentation functions obtained in this paper
(solid line) are compared at the scale $Q^2=10000 \ {\rm GeV}^2$ to 
those obtained by BKK ~\protect\cite{4r} (dotted line). 
}
\label{compbkk}
\end{figure}

In fig.~\ref{compbkk} we compare the result of our fit with the BKK fragmentation function
\cite{3r,4r} at $Q^2 = 10000 \ {\rm GeV}^2$. The two parametrizations are in reasonable
agreement for $z \ \lsim \ .6$~; however one must keep in mind that BKK studied pion and kaon cross sections, and
indirectly all charged cross sections (see below).  In fig.~\ref{ratbkk} we display ratios of
fragmentation functions on a linear scale in order to compare them more closely. \par

The $u$-quark and $b$-quark fragmentation functions are harder in BKK. The effect could be understood
partly as follows: the ratio of proton/pion cross section, inspired from data (fig. 5 of
\cite{alephr}), has been given the form $0.195-1.35(y-0.35)^2$ by BKK.
However, the high $y$ part of this form, not constrained by data,
is negative above 0.7 ; as a result their fit to all charged hadrons up to $y = 0.8$ will
have a tendency to enlarge some of the quark fragmentation functions in the large-$y$ region. On the
other hand, the $s$-quark fragmentation function is constrained by kaon production in BKK approach
and cannot be directly compared to our parametrization which does not distinguish between $s$-quark
and $d$-quark~; from Fig.~\ref{ratbkk} one can deduce that the ratio $(d+s)/(d+s)_{BKK}$ increases from 1.1 at
$y \simeq .1$ to 1.5 at $y \simeq .8$. The gluon fragmentation functions have different shapes but,
as already discussed, they are not very well constrained by $e^+e^-$ annihilation data. \par

The BKK collaboration also performed a direct fit of ALEPH charged hadron data and we can compare
the parametrizations given in the thesis of J. Binnewies \cite{bphd} with our results. The
ratios of the fragmentation functions are displayed in 
Fig. \ref{compbkk_charged}. The agreement between the
quark distributions is reasonable for $.1 \ \lsim \ z \ \lsim \ .6$. For $z > .6$, our
fragmentation functions are much larger than those of ref. \cite{bphd}. One reason for
this difference may come from the choice of the data. We choose to fit all LEP data. As a
result, our cross sections overshoot ALEPH data in the large-$z$ domain. On the other hand
Binnewies' parametrization slightly undershoots ALEPH data. Let us also notice that in
ref. \cite{bphd} ALEPH data is studied only in the range $z \leq .8$. Therefore we may
expect differences with our parametrizations for $z \ \gsim \ .8$. \par

Finally, one must also keep in mind that we used ``optimized'' cross sections
(expression (\ref{6e})) whereas BKK used the scale $M^2 = Q^2$. Moreover the bottom
threshold is set at $M^2 = 4 m_b^2$ by the BKK collaboration whereas we have $M^2 =
m_b^2$. This may introduce slight differences between the parametrizations. \par

The author of Ref. \cite{fra2} also compared 
his parametrization with the BKK one and observed disagreements for $z>~.5$ .
The behaviors of the
ratios shown in Fig. \ref{ratbkk} have similarities with those shown in Ref. \cite{fra2} for the
quark fragmentation functions. But one notes a strong disagreement 
for the gluon fragmentation function which is very small for $z > .5$ in Ref. \cite{fra2} compared to the 
BKK parametrization. Such a behavior seems to be in contradiction 
with the large-pt UA1 data discussed in the next section.
 where
an 83\% (54\%) contribution to the cross section is due to the
gluon fragmentation at $p_T=5 GeV/c$ ($p_T=21.9 GeV/c$).


\section{Analysis of large-p$_{\bot}$ production rate}
\hspace*{\parindent} The determination of the gluon fragmentation function $D_g(z, Q^2)$
from the inclusive $e^+e^-$ cross section is not very precise. The gluon contribution to the cross
section is a NLO effect, and, during the fitting procedure, we found that the gluon
parameters were quite sensitive to the functional form at $Q_0^2$ chosen for the quark
distributions. \par

A better constraint could come from the longitudinal cross section $d\sigma_L/dz$. But
data at large $z$, where we want to determine $D_g(z, Q^2)$, is rare and not accurate.
Therefore we turn to hadroproduction of large-$p_{\bot}$ hadron which constrains
the fragmentation function in the large-$z$ region ($z \simeq .7$). In hadronic collisions
at small $x_{\bot} = 2p_{\bot}/\sqrt{s}$, the contribution to the cross section involving
the gluon fragmentation function are large, and the theoretical predictions are quite
sensitive to them. \par

Data from the UA1 collaboration on charge hadrons at large $p_{\bot}$ (1 GeV $\lsim \
p_{\bot} \ \lsim$ 20 GeV) precisely cover this kinematical domain. For the theoretical
predictions, we use the code of ref. \cite{31r} which includes NLO corrections. The quark
and gluon distribution are from MRST-2 \cite{mrs99} and $\Lambda_{\overline{MS}} =$ 300 MeV.
The factorization and renormalization scales are set equal to $p_{\bot}/2$~; a choice which
is dictated by the stability of the cross section when the scales vary around
$p_{\bot}/2$ \cite{28r}. Our results for the region $p_{\bot} > 5$ GeV/c 
where we trust higher order QCD correction are compared with data in
fig.~\ref{mimi}. The slope and normalization are well reproduced, theory slightly
overshooting data. 
The  contribution to the cross section due to the gluon fragmentation
is important, going from 83\% to 54\%
from $p_T=5 GeV/c$ to $p_T=21.9 GeV/c$.
  
\par

We can try to improve the agreement between large-$p_{\bot}$ data and theory by
modifying the gluon parameters (\ref{8e}) within the allowed error range. With
$N_g = 5.25$ on the one hand, and $N_g = 1.1$ on the other hand, we obtain two
new sets of parameters and a $\chi^2$ value increased by one unity~: $\chi^2 =
202$.

\begin{equation}
\begin{array}{ll} 
 D_g(y,Q_0^2) &= 5.25  (1-y)^{1.87} y^{-0.03} \\
D_u(y,Q_0^2) &= \left ( 3.46 (1-y)^{2.26} -1.4(1-y)^{3.31} \right )
y^{-0.77}    \\
D_{d+s}(y,Q_0^2) &= \left ( 0.959 (1-y)^{0.95} + 8.86 (1-y)^{4.21} \right )
y^{-0.77} \\
D_c(y,Q_0^2) &= 4.07 (1-y)^{3.68} y^{-0.91}  \\
D_b(y,Q_0^2) &= 0.897 (1-y)^{3.72} y^{-1.7}  \\

\end{array}
\label{9.a}
\end{equation}

\begin{equation}
\begin{array}{ll} 
 D_g(y,Q_0^2) &= 1.1  (1-y)^{1.02} y^{-1.19} \\
D_u(y,Q_0^2) &= \left ( 3.32 (1-y)^{2.24} -0.5(1-y)^{3.39} \right )
y^{-0.62}    \\
D_{d+s}(y,Q_0^2) &= \left ( 0.93 (1-y)^{0.94} - 0.5 (1-y)^{4.54} \right )
y^{-0.62} \\
D_c(y,Q_0^2) &= 6.37 (1-y)^{4.06} y^{-0.68}  \\
D_b(y,Q_0^2) &= 0.897 (1-y)^{3.72} y^{-1.7}  \\

\end{array} 
\label{9.b}
\end{equation}

The corresponding fragmentation functions are compared to those obtained from
the input (\ref{8e}) in Fig. \ref{mar18} and \ref{apr20} 
for $Q^2 = 10 000$~GeV$^{2}$. We note a
change in the gluon shape which is important at large $z$. Unfortunately the
large-$p_{\bot}$ cross section tests the fragmentation functions in the domain
$z \simeq .7$ where the change is small, and we do not expect a substantial
modification of the large-$p_{\bot}$ cross section. This point is verified in
Fig.~\ref{mimi_mar18} and \ref{mimi_apr20} which shows only a 
very slight improvement of the ratio data/theory.
Therefore we conclude that the 3 sets of parameters (\ref{8e}), (\ref {9.a}) and
(\ref{9.b}) are compatible with UA1 data.

\begin{figure}[htb]
\begin{center}
\epsfig{file=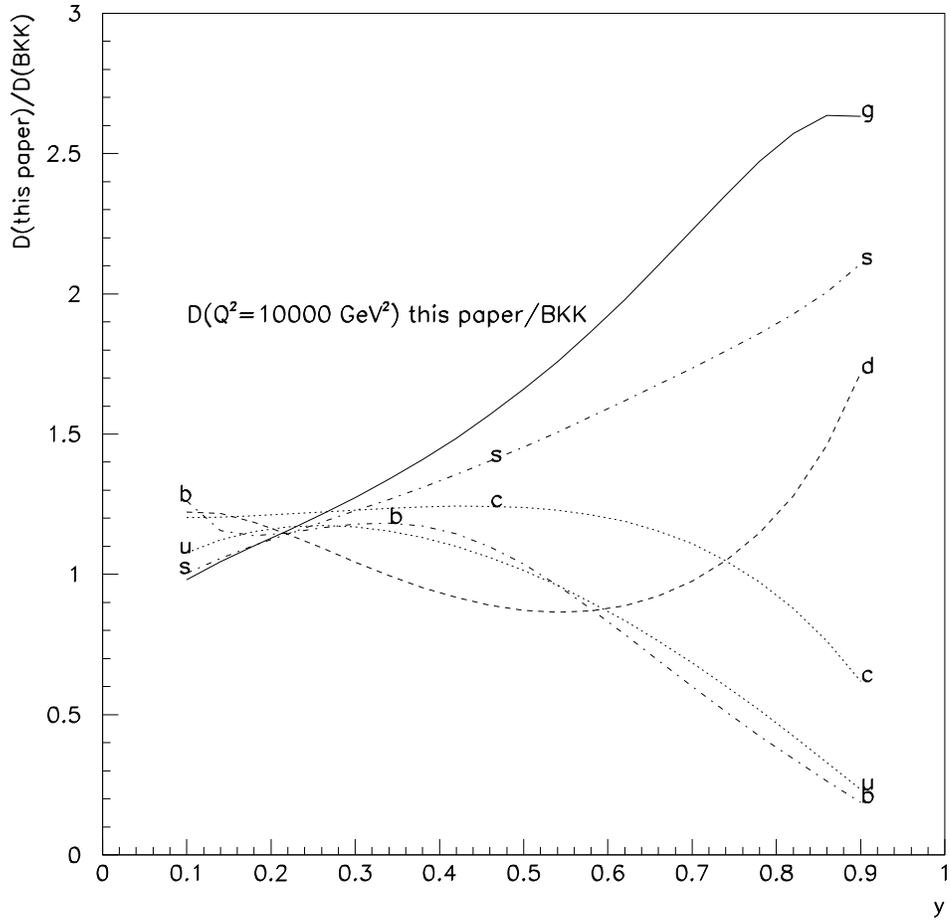,height=14.cm,width=14.cm}
\end{center}
\caption{Ratios of the parton fragmentation functions obtained in this paper (formula \ref{8e}) to 
those obtained by BKK ~\protect\cite{4r}. The scale is $Q^2=10000 \ {\rm GeV}^2$. 
}
\label{ratbkk}
\end{figure}

\begin{figure}[htb]
\begin{center}
\epsfig{file=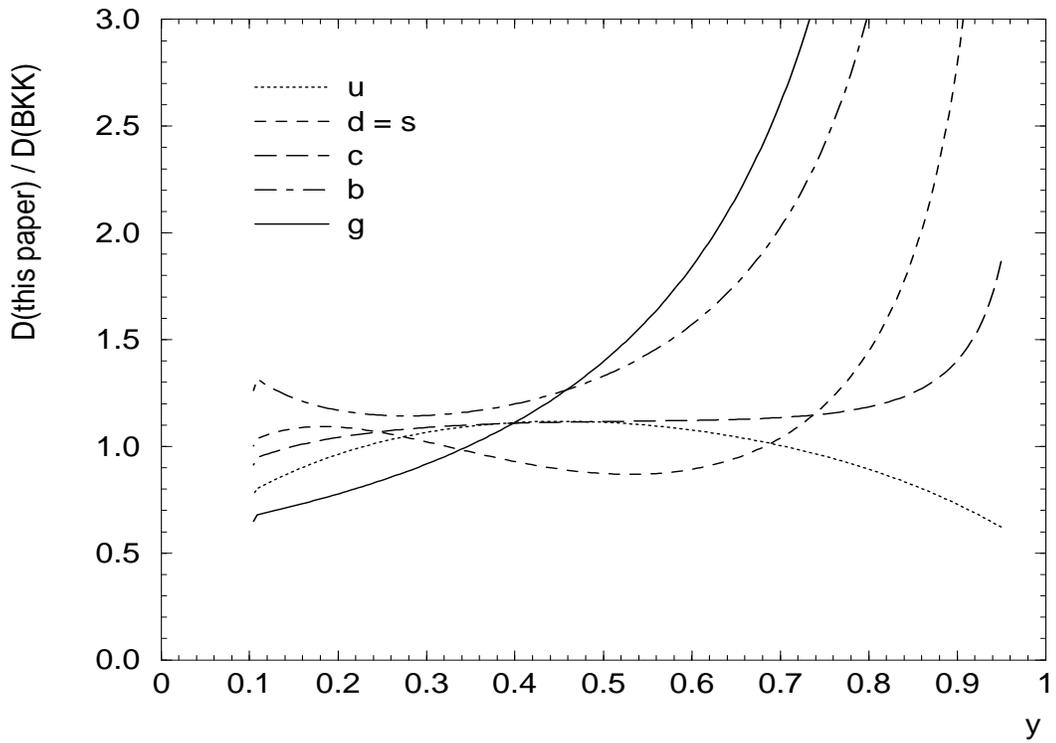,height=10.cm,width=14.cm.}
\end{center}
\caption{Quark fragmentation functions obtained in this paper
(solid line) are compared at the scale $Q^2=10000 \ {\rm GeV}^2$ to 
those obtain in ~\protect\cite{bphd} using all charged  (dotted line).
}
\label{compbkk_charged}
\end{figure}  

\begin{figure}[htb]
\begin{center}
\epsfig{file=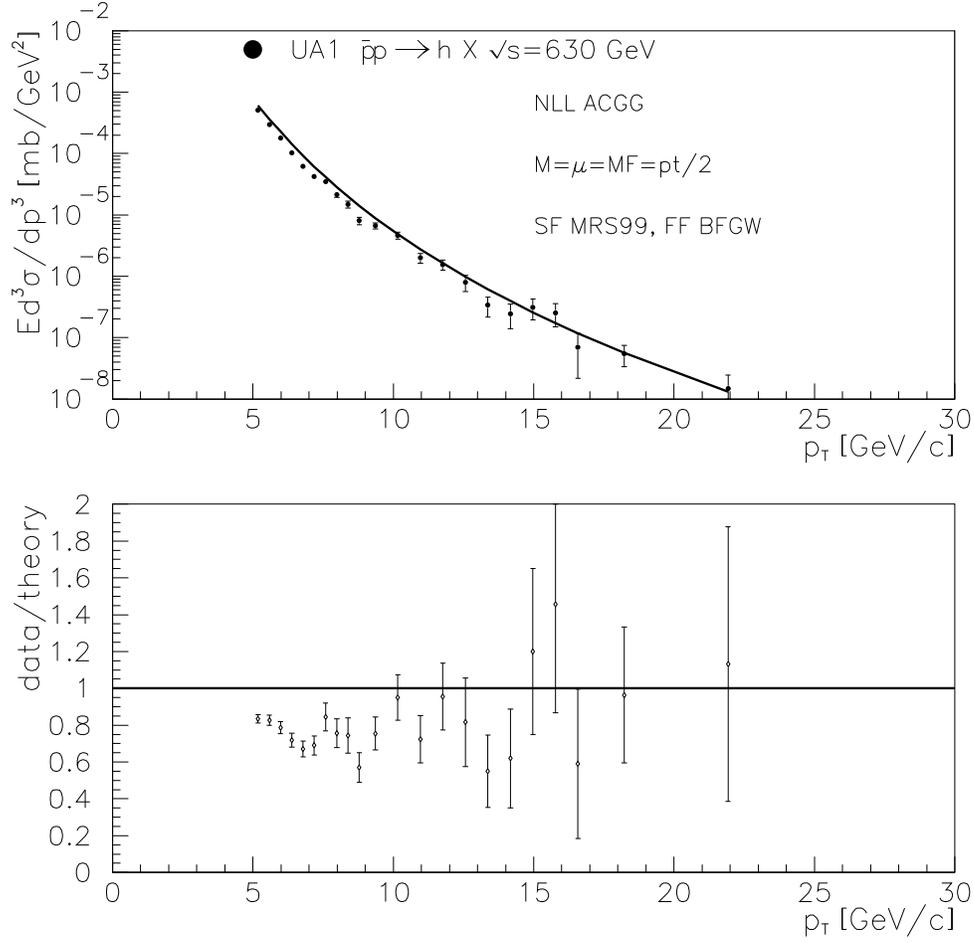,width=14cm}
\end{center}
\caption{NLO inclusive charged particle production in $p\bar{p}$ collisions
at $\sqrt{s}=630$ GeV for $\mu=M=M_F=p_T/2$ with fragmentation functions obtained 
here (formula \ref{8e}) and structure functions MRS99-2 ~\protect\cite{mrs99} compared to
data of the UA1 collaboration~\protect\cite{11r}. 
}
\label{mimi}
\end{figure}

\begin{figure}[htb]
\begin{center}
\epsfig{file=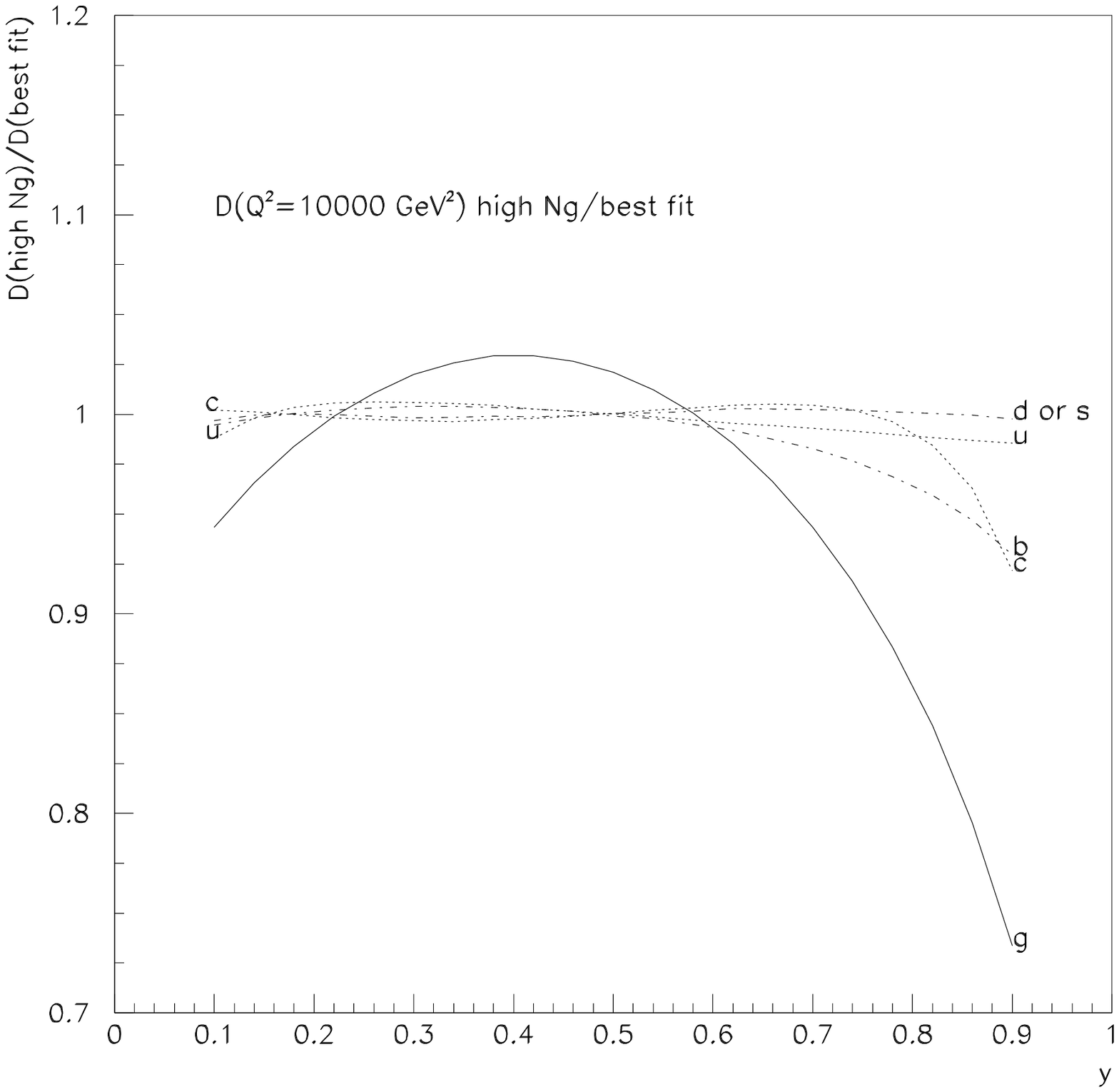,height=14.cm,width=14.cm}
\end{center}
\caption{Ratio of the fragmentation functions given by (\ref{9.a})
to those of the best fit (\ref{8e}) at the scale $Q^2=10000 \ {\rm GeV}^2$. }
\label{mar18}
\end{figure}

\begin{figure}[htb]
\begin{center}
\epsfig{file=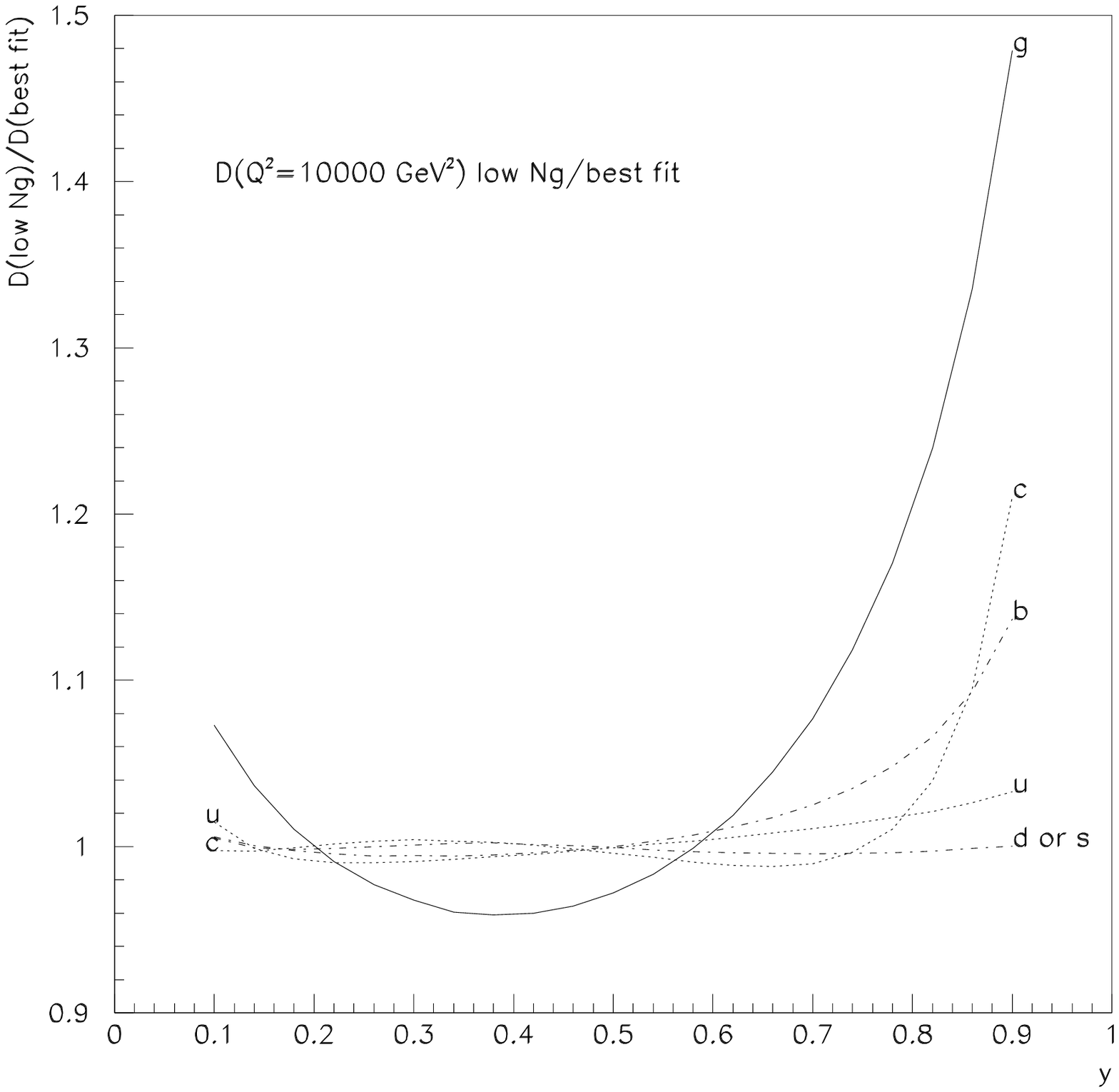,height=14.cm,width=14.cm}
\end{center}
\caption{Ratio of the fragmentation functions given by (\ref{9.b})
to those of the best fit (\ref{8e}) at the scale $Q^2=10000 \ {\rm GeV}^2$. }
\label{apr20}
\end{figure}

\begin{figure}[htb]
\begin{center}
\epsfig{file=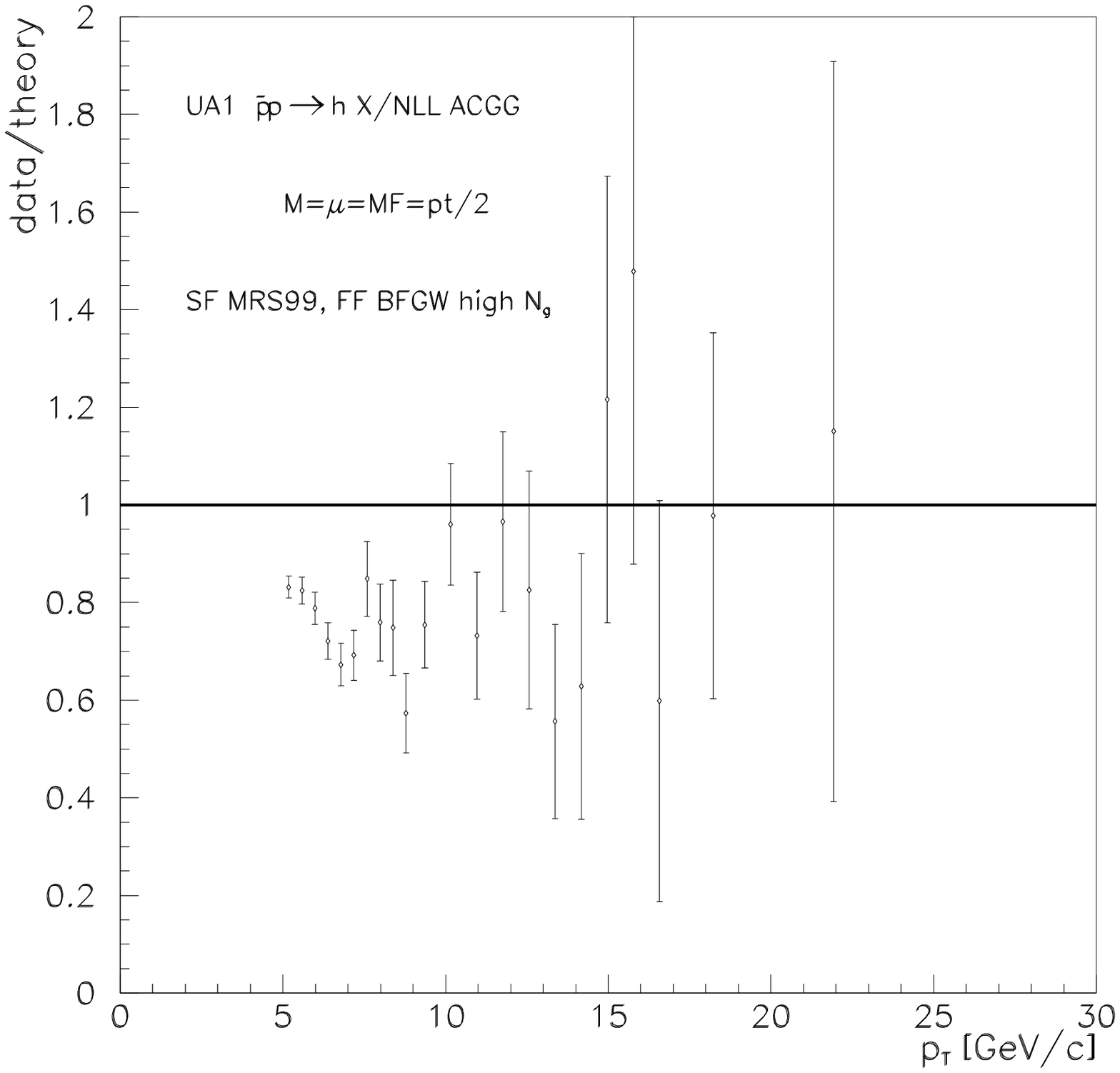,width=14cm}
\end{center}
\caption{NLO inclusive charged particle production in $p\bar{p}$ collisions
at $\sqrt{s}=630$ GeV for $\mu=M=M_F=p_T/2$ with the set of fragmentation functions obtained here 
(\protect\ref{9.a})
by allowing a $\Delta\chi^2$ of 1 to the best fit and structure functions MRS99-2 ~\protect\cite{mrs99} compared to
data of the UA1 collaboration~\protect\cite{11r}. 
}
\label{mimi_mar18}
\end{figure}

\begin{figure}[htb]
\begin{center}
\epsfig{file=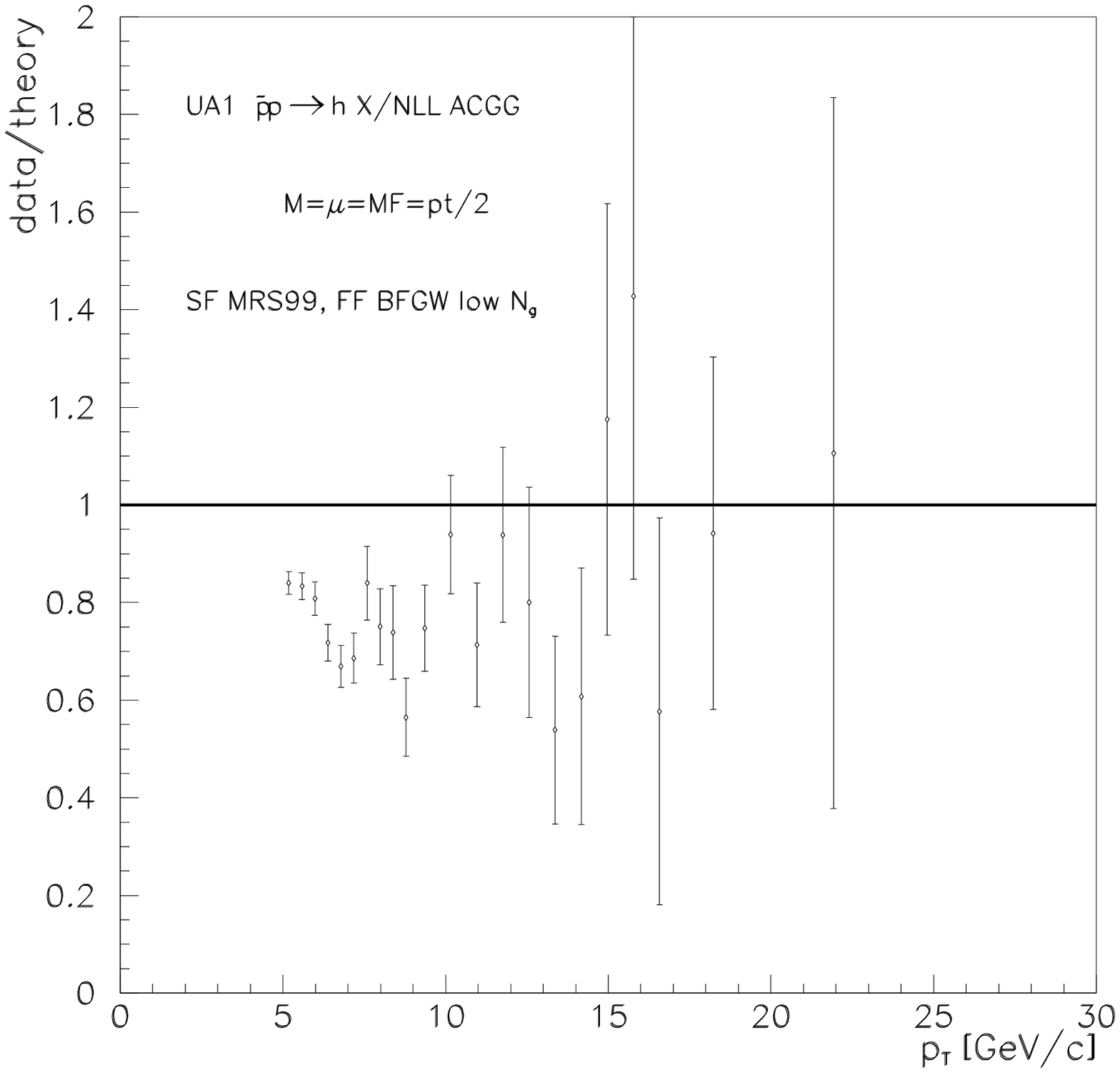,width=14cm}
\end{center}
\caption{NLO inclusive charged particle production in $p\bar{p}$ collisions
at $\sqrt{s}=630$ GeV for $\mu=M=M_F=p_T/2$ with the set of fragmentation functions obtained here (\protect\ref{9.b})
by allowing a $\Delta\chi^2$ of 1 to the best fit and structure functions MRS99-2 ~\protect\cite{mrs99} compared to
data of the UA1 collaboration~\protect\cite{11r}. 
}
\label{mimi_apr20}
\end{figure}

\section{Conclusion}
\hspace{\parindent} In this paper we have performed a NLO analysis of $e^+e^-$ annihilation into
charged hadrons data in order to determine a new set of fragmentation functions. Although many LEP
data are now available, including samples with enhanced content in heavy quarks, we find that it is
impossible to obtain a convergent fit when using three (or five for the $u$-quark) free parameters to
characterize the input shapes of the quark and gluon fragmentation functions. There are strong
correlations between the parameters and we do not obtain a positive definite error matrix. 
 \par

Fixing some of the strongly correlated parameters to values obtained in a fit characterized by a good
$\chi^2$, we succeed in obtaining a positive definite error matrix. The overall
agreement with data is quite good. Comparing our parametrizations of the
fragmentation functions with those obtained by the BKK collaboration \cite{4r},
we note several important discrepancies, especially in the large-$z$ domain.
These discrepancies should come from the assumption made by the BKK collaboration
on the large-$z$ behavior of the charged hadron cross sections. A better
agreement is reached with the results of ref. \cite{bphd} in which charged cross
sections are also studied. But in the large-$z$ region, one again notes important
discrepancies. One explanation of the discrepancy could be the fact that we use
sets of data extending to larger $z$-values than those fitted in \cite{bphd}. \par

This point also emphasizes the necessity to carefully treat the large-$z$ region and to resum the
large logarithms in the theoretical expressions. In this work this is done through the optimized
scales which should also be used in the calculation of other reactions making use of the present
parametrization. \par

We also test the gluon fragmentation function in large-$p_{\bot}$ hadronic
collisions. We find that 3 sets of fragmentation functions allowed by $e^+e^-$
data within one standard deviation and differing in the gluon shape lead to very
similar predictions for large-$p_{\bot}$ cross sections at the $Sp\bar{p}S$
energy. The overall agreement with UA1 data is quite good.

\section{Appendix}
 
 Uncertainties on the parameters of the fit (\ref{8e}) can be roughly 
estimated  with the error matrix using the curvature at the minimum
and assuming a parabolic shape \cite{mini}, \cite{eadie}:
the results are given on Table \ref{table1} where the parabolic errors quoted
take into account the effects due to parameter correlations.
These errors should be taken as lower limits on the errors,
as some parameters, which may be correlated to the ones estimated,
have been fixed in the fitting procedure. The results of a full
MINOS analysis (non-parabolic chisquare) \cite{mini,eadie}, 
following the $\chi^2$
out of the minimum and finding where it corresponds
to  $\Delta_{\chi^2}=1$ is given on Table \ref{table2}.

\begin{table*}[htb]
\caption{Errors on the parameters and correlation}
\begin{scriptsize}
\begin{tabular}{|c|c|c|c|c|c|c|c|c|c|c|c|c|c|}
\hline
par.&error& \multicolumn{12}{c}{correlation with} \\
& &$\alpha_g$&$\beta_g$ &$N_{d+s}$ &$\alpha_u$ & $\beta_{d+s}$ & $N_u$ & $\bar{N}_u$ 
& $\beta_u$ &$\bar{N}_{d+s}$ &$N_c$& $\alpha_c$ & $\beta_c$ \\
\hline
$N_g $& 1.49 & 0.98& 0.97& 0.11& -0.66 &0.10 &-0.003 &-0.16 &-0.03 &-0.71
&-0.6 & -0.6 & -0.5 \\
$\alpha_g $& 0.53 & & 0.90 & 0.04 &-0.75 &0.05 &0.01 &-0.19 &-0.02 &-0.78
&-0.6 & -0.6 & -0.5 \\
$\beta_g$ & 0.40 & & & 0.22 &-0.52 &0.18 &-0.02 &-0.12 &-0.05 &-0.58
&-0.5 & -0.5 & -0.5 \\
$N_{d+s}$& 0.16 & & & & 0.20 &0.97&-0.57 &0.57&-0.49 &0.10&
0.08 & 0.12 & -0.03\\
$\alpha_u$& 0.09 & & & & &0.16&-0.35& 0.55&-0.32 &0.98&
0.29 & 0.36&0.14\\
$\beta_{d+s}$& 0.09 & & & & & &-0.62 &0.58 &-0.57& 0.08 & 
0.09 & 0.12 & 0.01\\
$N_u$ & 1.4 & & & & & & &-0.97& 0.97 &-0.30 &
 0.0 & -0.04 & 0.07 \\
$\bar{N}_u$ & 2.0 & & & & & & & &-0.92 & 0.49 &
0.05 & 0.11 & -0.05 \\
$\beta_u$& 0.25 & & & & & & & & &-0.30
& -0.04 & -0.06 & 0.01\\
$\bar{N}_{d+s}$ & 1.8 & & & & & & & & & &
0.30 & 0.35 & 0.16 \\
$N_c$&1.77       & & & & & & & & & & & 0.99 & 0.95\\
$\alpha_c$& 0.17& & & & & & & & & & & & 0.90\\
$\beta_c$&0.33 & & & & & & & & & & & & \\
\hline
\end{tabular}
\end{scriptsize}
\label{table1}
\end{table*}

\begin{table*}[htb]
\caption{Uncertainties on the parameters}
\begin{tabular}{|c|c|c|c|}
\hline
parameter &parabolic error& negative error & positive error\\
\hline
$N_g = 2.12$& 1.49 & -1.02 & +3.1 \\
$\alpha_g = -0.71$& 0.53& -0.51 & +0.70 \\
$\beta_g = 1.37$ & 0.40 & -0.38 & +0.51 \\
$N_{d+s} = 0.948 $& 0.157 & -0.15 & +0.17\\
$\alpha_u = -0.684$& 0.099 & -0.11 & +0.098\\
$\beta_{d+s} = 0.944$& 0.091 & -0.090 & +0.096\\
$N_u = 3.435$ & 1.40& -1.56& +1.35 \\
$\bar{N}_u = -0.956$ & 2.01 & -1.94 &+2.19 \\
$\beta_u = 2.26$& 0.25 & -0.33 &+0.22 \\
$\bar{N}_{d+s} = 10.43$ & 1.83 & -1.95& +1.93\\
$N_c=5.15$&1.77&-1.54&+2.26 \\
$\alpha_c=-0.789$& 0.177&-0.184 &+0.186\\
$\beta_c=3.87$&0.33&-0.33&+0.35\\
\hline
\end{tabular}
\label{table2}
\end{table*}

\end{document}